\documentstyle[12pt]{article}
\begin{document}

\title{A New Approach to Flavor Symmetry and an Extended Naturalness Principle}

\author{{\bf S.M. Barr} \\
Department of Physics and Astronomy \\ University of Delaware \\
Newark, Delaware 19716}

\date{\today}

\maketitle

\begin{abstract}
A class of non-supersymmetric extensions of the Standard Model is
proposed in which there is a multiplicity of light scalar doublets
in a multiplet of a non-abelian family group with the Standard Model
Higgs doublet. Anthropic tuning makes the latter light, and
consequently the other scalar doublets remain light because of the
family symmetry. The family symmetry greatly constrains the pattern
of FCNC and $p$ decay operators coming from scalar-exchange. Such
models show that useful constraints on model-building can come from
an extended naturalness principle when the electroweak scale is
anthropically tuned.
\end{abstract}

\newpage

\section{Introduction}

In models with low-energy supersymmetry (SUSY), the supersymmetry
allows the existence at low energy of many scalar fields in a way
consistent with the old ``naturalness principle" \cite{naturalness},
because of the non-renormalization theorems. Supersymmetry it also
allows great leeway in the structure of the superpotential without
violating the naturalness principle. The consequent flexibility in
model building is both an advantage and a disadvantage. One
disadvantage is that a huge number of terms are permitted whose
coefficients are not constrained by any principle (unless somewhat
ad hoc symmetries are imposed). Many of these terms violate baryon
number, flavor or CP. Thus, the very flexibility of low-scale SUSY
seems to undo several of the greatest successes of the Standard
Model, which were due to the fact that the symmetries of the
Standard Model greatly restrict the couplings that can be written
down.

If low-energy supersymmetry is abandoned there is one large price to
pay, namely the fine-tuning of the mass-squared parameter ($\mu^2$)
of the Standard Model (SM) Higgs doublet. That is not necessarily a
disaster, however, as such a tuning might be anthropically accounted
for \cite{abds, bk, sweinberg}, and there is a corresponding gain,
namely more constrained model-building. Without low-energy
supersymmetry, the possibilities for new particles at or near the
electroweak scale are more limited and their small masses must
either be dynamically generated or protected by conventional (i.e.
non-SUSY) symmetries, such as chiral symmetry and gauge invariance,
which may place strong conditions on their interactions. (One sees
this, of course, in technicolor models, which are so highly
constrained that it is difficult to construct realistic models based
on the idea.)

In this paper, we consider models without low-energy supersymmetry,
and propose an extension of the old naturalness principle, which we
call the ``extended naturalness principle". According to this
principle, there should be no parameters in a model that are very
small (or otherwise take very special values) unless this has {\it
either} a conventional ``natural" explanation in terms of symmetry
principles and dynamical mechanisms, {\it or} can be accounted for
by ``anthropic tuning." We will see, by considering a certain class
of models, that this extended naturalness principle can greatly
constrain model building (as the old naturalness principle did) and
lead to interesting and testable scenarios.

The kind of model that will be considered in this paper is
characterized by having a multiplicity of light scalar doublets
(instead of just one, as in the Standard Model). One reason for
considering such a possibility is that the presence of several (six
or even five) light scalar doublets can lead to satisfactory
unification of gauge couplings without supersymmetry
\cite{sixdoublets, sww}. But to make a multiplicity of scalar
doublets light in a way consistent with the extended naturalness
principle requires new symmetries. The basic idea assumed here is
that these symmetries are family symmetries. The ``extra" light
scalar doublets will be assumed to be in a multiplet of a
non-abelian group $G_F$ with the SM Higgs doublet. That means that
when the mass of the SM Higgs doublet is made small by anthropic
tuning, so will the entire multiplet of scalar doublets.

The idea that the tiny (and negative) mass-squared of the SM Higgs
may be the result of anthropic tuning in what is now called a
``multiverse" was proposed in \cite{abds}. Some reasons to regard
this possibility as plausible have been stated by S. Weinberg: ``If
the electroweak scale is anthropically fixed, then we can give up
the decades long search for a natural solution to the hierarchy
problem. This is a very attractive prospect, because none of the
`natural' solutions that have been proposed, such as technicolor or
low energy supersymmetry, were ever free of difficulties.  In
particular, giving up low energy supersymmetry can restore some of
the most attractive features of the non-supersymmetric standard
model: automatic conservation of baryon and lepton number in
interactions up to dimension 5 and 4 respectively; natural
conservation of flavors in neutral currents; and a small neutron
electric dipole moment" \cite{sweinberg}.

If the SM Higgs doublet is in a non-trivial multiplet of $G_F$ and
gives mass to the known quarks and leptons through renormalizable
$d=4$ Yukawa operators, then the known quarks and leptons would also
have to be in non-trivial $G_F$ multiplets. In that case, $G_F$
would be a ``family symmetry". This is an attractive possibility as
there would be a single explanation for both the multiplicity of the
light quarks and leptons and the multiplicity of the light scalar
doublets, namely that they were multiplets of a family symmetry.
(Moreover, having ``families" of both the fermions and the scalars
seems less unbalanced than the Standard Model, where there are large
numbers of quarks and leptons, but only one scalar multiplet.) The
most obvious possibilities for $G_F$, given the fact that there are
three light families of quarks and leptons, are $SU(3)$ and $SO(3)$,
with the light families being in a ${\bf 3}$ in either case. If $G_F
= SU(3)$, there could be six light scalar doublets either in ${\bf
3} + {\bf 3}$ or in ${\bf 6}$. If $G_F = SO(3)$, there could be six
light scalar doublets in ${\bf 3} + {\bf 3}$ or five of them in
${\bf 5}$. We shall consider these possibilities briefly later, and
shall discuss a toy model based on $SO(3)_F$ in section 3, but we
shall find it easier to construct a realistic model based on
$SO(4)_F$, which will be discussed in section 4.

The ``extra" light scalar doublets must have masses of at least
several TeV to avoid excessive flavor violation in neutral current
processes. Thus, there must be splitting of the $G_F$ multiplet of
scalar doublets by an amount at least an order of magnitude greater
than the mass of the SM Higgs doublet.  The idea is that the {\it
overall} mass-squared of the $G_F$ multiplet of scalar doublets
(i.e. the $G_F$-invariant part) ``scans" among the ``domains" or
``subuniverses" of the universe and is anthropically set to have a
value (in our ``subuniverse" or ``domain") such that the {\it
lightest} member of the $G_F$ multiplet (i.e. the SM Higgs doublet)
has a mass-squared that is negative and of order -(100 GeV)$^2$.
That means that all the other members of the $G_F$ multiplet will
have masses-squared that are of order $M_F^2$, where $M_F^2 \gg 1$
TeV is the scale of splitting of the multiplet. (Of course, it must
be only the lightest scalar doublet whose mass-squared is pushed
negative, since if more than one is, the one with most negative
mass-squared will get a VEV of order $M_F$, which is not
anthropically viable \cite{abds}.) The breaking of $G_F$ can be
dynamical, and thus $M_F$ can naturally be much smaller than the
unification scale, and in particular in the TeV range. Figure 1
illustrates the idea schematically.

\newpage

\begin{picture}(360,180)
\put(180,75){\line(0,1){10}} \put(180,95){\line(0,1){10}}
\put(180,115){\line(0,1){10}} \put(180,1355){\line(0,1){10}}
\put(160,30){\line(1,0){60}} \thicklines
\put(160,27){\line(1,0){40}} \put(160,33){\line(1,0){40}}
\put(160,37){\line(1,0){40}} \put(160,41){\line(1,0){40}}
\put(160,45){\line(1,0){40}} \put(160,175){\line(1,0){40}}
\put(155,37){\vector(0,1){10}} \put(155,37){\vector(0,-1){10}}
\put(112,33){$O(M_F^2)$} \put(122,171){$M_{unif}^2$}
\put(225,26){$m^2=0$} \put(160,0){{\bf Fig. 1}}
\end{picture}

\vspace{0.5cm}

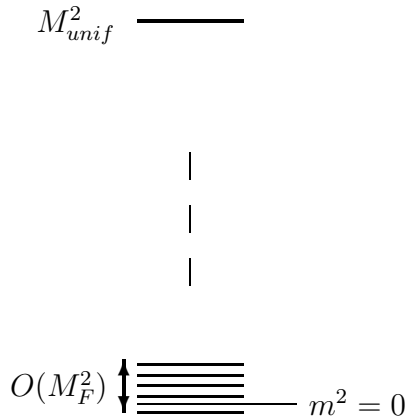
\noindent {\bf Fig. 1.} The overall ($G_F$-invariant) mass-squared
of the $G_F$ multiplet of scalar doublets is tuned so the lightest
($H_{SM}$) has small negative mass-squared. The others then have
positive mass-squared from $G_F$ breaking.

\vspace{0.5cm}

An obvious issue if the unification scale is around $10^{14}$ GeV
(as in the non-SUSY SM with six scalar doublets) is rapid proton
decay. In \cite{sww} this was avoided by assuming the unified group
to be the ``trinification" group $SU(3)_c \times SU(3)_L \times
SU(3)_R \times S_3$, where the $S_3$ cyclically permutes the three
$SU(3)$ factors. We shall use the same group. (For an outline of the
trinification scheme assumed in this paper, see Appendix A.) This
eliminates proton decay mediated by gauge boson exchange. However,
since some of the extra light scalar doublets in our models shall
necessarily couple to light quarks and leptons with couplings of
order 1 (as will be seen), proton decay by the exchange of their
superheavy colored partners becomes an issue. It is shown in
Appendix B that there exist terms in the scalar potential that can
give the dangerous colored scalars masses large enough to avoid
excessive proton decay. The proton decay that does exist, however,
will have characteristic branching ratios determined by the symmetry
$G_F$, as will be discussed later.

Another possibility besides trinification is unification based on a
simple group, such as $SU(5)$ or $SO(10)$, broken by orbifold
compactification \cite{kawamura}. If the light quarks and leptons
live on a brane where there is only the SM gauge group, proton decay
can be suppressed, since the gauge bosons of $G_{GUT}/G_{SM}$ will
vanish on the SM brane by the orbifold boundary conditions, and
similarly for the colored scalars that would mediate proton decay.
On the other hand, gauge kinetic terms in the 4D lagrangian on the
SM brane would not respect $G_{GUT}$ and would affect the gauge
unification. If these effects were small for some reason, then
unification could still take place with 5 or 6 light scalar
doublets. We shall not pursue this possibility here, but will
henceforth assume the unified group to be $SU(3)_c \times SU(3)_L
\times SU(3)_R \times S_3$, which will be denoted $G_U$ for short.
The irreducible multiplet that contains a family, namely $(3,
\overline{3}, 1) + ( \overline{3}, 1, 3) + ( 1, 3, \overline{3})$
will be denoted henceforth as $F$.

In the later sections it will be seen that the inter-related set of
assumptions we have made (no low-energy SUSY, unification of gauge
couplings by a multiplicity of light scalar doublets whose mass is
related to that of the SM Higgs by a family symmetry) leads to
tightly constrained possibilities for model building and interesting
and novel phonomenological consequences. While anthropic tuning in
general, and of the electroweak scale in particular, cannot be
tested directly, this shows that when it is combined with the
extended naturalness principle and other attractive assumptions
predictive schemes can result. The rest of the paper will be
organized as follows. In section 2, some basic points about the
anthropic tuning of the electroweak scale will be discussed. In
section 3, a simple $SO(3)_F$ toy model will be presented and its
inadequacies pointed out. In section 4, a realistic model based on
$SO(4)_F$ will be described and analyzed.  Some details having to do
with the trinification group and its breaking will be discussed in
the Appendices.

\section{Anthropic Tuning of the Electroweak Scale}

The idea that the electroweak scale might be anthropically
determined in the context of what is now called the multiverse
scenario was proposed in \cite{abds}. There it was noted that the
cosmological constant $\Lambda$ ($\approx 10^{-120} M_{P \ell}^4$)
and the Higgs mass parameter of the Standard Model $\mu^2$ ($\approx
10^{-34} M_{P \ell}^2$) are the two smallest parameters in our
current theory (in natural units) as well as the ones whose
smallness has proven most difficult to explain in conventional ways.
(The next smallest parameter $\overline{\theta}$ is only bounded to
be less than about $10^{-10}$, and several viable ``natural"
explanations exist for its smallness \cite{strongCP}. Many
technically natural symmetry explanations have also been proposed
for the small quark and lepton Yukawa couplings of the lighter
families. And the explanation of the smallness of
$\Lambda_{QCD}/M_{P \ell}$ in terms of the logarithmic running of
$\alpha_s$ and dimensional transmutation is perfectly ``natural".)
It was therefore suggested in \cite{abds} that $\Lambda$ and $\mu^2$
are the most plausible candidates for anthropic tuning. This was
also suggested by S. Weinberg \cite{sweinberg}: ``The most
optimistic hypothesis is that the only constants that scan are the
few whose dimensionality is a positive power of mass: the vacuum
energy and whatever mass or masses set the scale of electroweak
symmetry breaking."

The reason that the magnitude of $\mu^2$ must be at or lower than
(100 GeV)$^2$ for the evolution of life to be at all plausible is
that larger (negative) values of $\mu^2$ lead to larger $v$ ($\equiv
\langle H_{SM} \rangle$), and therefore larger quark masses. This
leads to larger pion masses and shorter range of the nucleon-nucleon
potential. This makes nuclei more unstable, beginning with the
all-important deuteron, which becomes unbound if $v$ is larger than
its observed value by a factor of about 1.4 to 2.7 \cite{abds}. The
case of positive $\mu^2$ require separate arguments, given in
\cite{abds} (for $\Lambda =0$) and \cite{bk} (for $\Lambda \neq 0$).

An extension of the analysis of \cite{abds} to the case of two
doublets $H_u$ and $H_d$, which transform under the Standard Model
gauge group as $(1,2 + \frac{1}{2})$ and $(1,2,- \frac{1}{2})$
respectively (and which couple respectively to the up quarks and to
the down quarks and charged leptons) was give in \cite{bk}. This is
relevant to the present work, since such a pair of doublets exists
in many unified schemes, including trinification. With two doublets,
there is a $2 \times 2$ mass-squared matrix that can be written as

\begin{equation}
\left( H_u, H_d^* \right) \left( \begin{array}{cc} M^2_u & \Delta^2
\\ \Delta^{2*} & M_d^2 \end{array} \right) \left( \begin{array}{c}
H_u^* \\ H_d \end{array} \right).
\end{equation}

\noindent Only one fine-tuning is required to make the SM doublet
light, namely of the determinant of the mass-squared matrix, with
all the elements of the matrix remaining of the unification scale.
The light doublet would be

\begin{equation}
\begin{array}{l}
H_{SM} = \cos \theta_H H_u + e^{i \alpha} \sin \theta_H H^*_d, \\ \\
\tan 2 \theta_H \equiv 2 |\Delta^2|/(M_u^2 - M_d^2) = O(1), \\ \\
\alpha \equiv \arg \Delta^2.
\end{array}
\end{equation}

\noindent The other mass eigenstate, which remains superheavy, is
$H_h = - e^{-i \alpha} \sin \theta_H H_u + \cos \theta_H H^*_d$. In
typical unified models, where $H_u$ couples to up quarks and $H_d$
couples to down quarks and charged leptons, this will be reflected
in the Yukawa couplings of the SM Higgs field at low energy, which
will be proportional to $\cos \theta_H$ for the up quarks and
proportional to $\sin \theta_H$ for the down quarks and charged
leptons. If at least two independent combinations of the parameters
$M_u^2$, $M_d^2$, and $\Delta^2$ ``scan" among the domains, then
both $\mu^2$ (the mass-squared parameter of the SM Higgs) and $\tan
\theta_H$ scan. As shown in \cite{bk}, not only do anthropic
considerations set $\mu^2$ to be near -(100 GeV)$^2$, but they also
set $\tan \theta_H$ such that the $d$-quark mass is comparable to,
but slightly larger than, the $u$-quark mass, as indeed observed.
(See also \cite{hogan}.)

In our case, the scalar doublets are in a representation $R$ of the
family group $G_F$. Denoting them $H_u^A$ and $H_d^A$, where $A = 1,
... , R$, one has

\begin{equation}
\left( H_u^A, H_{dA}^* \right) \left( \begin{array}{cc} M^2_u \;
\delta_A^B & (\Delta^2)_{AB} \\
\\ (\Delta^{2*})^{AB} & M_d^2 \; \delta^A_B \end{array} \right)
\left( \begin{array}{c} H_{uB}^* \\ H_d^B \end{array} \right).
\end{equation}

\noindent Note that if $G_F = SU(N)$ and $R$ is an $N$ or other
complex representation, then $(\Delta^2)_{AB}$ breaks $G_F$ and must
be near the electroweak scale. In that case, a single tuning to make
the determinant of the matrix small is not sufficient. Such a tuning
would make {\it either} $M_u^2$ {\it or} $M_d^2$ to be of order the
weak scale, which would result in the light Higgs doublets being
almost purely of the type $H_u$ or of the type $H_d$. That would not
give electroweak-scale masses to all the quarks and leptons. In
order for all the light quarks and leptons to obtain realistic
masses, {\it both} $M^2_u$ {\it and} $M_d^2$ would separately have
to be tuned to be of order the weak scale.

On the other hand, if $G_F = SO(N)$ and $R$ is an $N$ or other real
representation, then the matrix can have the form

\begin{equation}
\left( H_u^A, H_d^{*A} \right) \left( \begin{array}{cc} M^2_u &
\Delta^2 \\
\\ \Delta^{2*} & M_d^2 \end{array} \right) \; \delta^{AB}.
\left( \begin{array}{c} H_u^{*B} \\ H_d^B \end{array} \right);
\end{equation}

\noindent In this case, all the elements of the mass matrix can be
of order the unification scale and a {\it single} tuning of the
determinant is enough to make a multiplet of scalar doublets light
that is a mixture of both $H_u$ and $H_d$ type. Specifically, what
will be made light from a single tuning is an $R$-multiplet of
scalar doublets $H^A \equiv (\cos \theta_H H_u^A + e^{i \alpha} \sin
\theta_H H^{*A}_d)$, where $A = 1, ... , R$.

Not only do models with $G_F = SO(N)$ require fewer fine-tunings, in
this sense, but the problem of canceling $G_F$ anomalies does not
arise. We shall therefore consider only models with orthogonal
family groups.

\section{An $SO(3)_F$ Toy Model}

Several key characteristics of the kind of model being proposed in
this paper can be understood in a simple toy model with $G_F =
SO(3)$. (A realistic model with $G_F = SO(4)$ will be discussed in
section 4.) The gauge group of the model is

\begin{equation}
\begin{array}{ll}
G = & G_U \times G_F \times G_{DSB} \\ \\
& G_U = SU(3)_c \times SU(3)_L \times SU(3)_R \times S_3 \\ \\
& G_F = SO(3)_F \\ \\
& G_{DSB} = SU(N)_{DSB} \end{array}
\end{equation}

\noindent The confining group $SU(N)_{DSB}$ plays the role of
dynamically breaking the family group $SO(3)_F$. The three pieces of
the $G_F$ multiplet $F$, namely $(3, \overline{3}, 1)$, $(
\overline{3}, 1, 3)$, and $(1, 3, \overline{3})$, will be denoted by
the subscripts $q$, $\overline{q}$, and $\ell$, respectively. The
three families are in a $(F, 3, 1)$ of $G_U \times G_F \times
G_{DSB}$ that will be denoted $\psi^i = \psi^i_q +
\psi^i_{\overline{q}} + \psi^i_{\ell}$, where $i$ is a $SO(3)_F$
vector index. The light scalar doublets are in a $(F, 5, 1)$ of $G_U
\times G_F \times G_{DSB}$, which will be denoted $\Phi^{(ij)} =
\Phi^{(ij)}_q + \Phi^{(ij)}_{\overline{q}} + \Phi^{(ij)}_{\ell}$.
Note that $\Phi^{(ij)}_{\ell}$ is a rank-2 symmetric traceless
tensor of $SO(3)_F$ and therefore contains 5 doublets $H^{(ij)}_u$
and 5 doublets $H^{(ij)}_d$. The Yukawa terms of the quarks and
leptons are therefore just

\begin{equation}
{\cal L}_{Yukawa} = y \left[ \left( \psi^i_q \;
\psi^j_{\overline{q}} \right) \Phi^{(ij)}_{\ell} + cyclic \right] +
y' \left[ \left( \psi^i_{\ell} \; \psi^j_{\ell} \right)
\Phi^{(ij)}_{\ell} + cyclic \right]
\end{equation}

\noindent The expression ``+ {\it cyclic}" refers to the $S_3$
permutations of the three $SU(3)$ trinification groups that take $q
\longrightarrow \overline{q} \longrightarrow \ell \longrightarrow
q$.

The $\Phi^{(ij)}_{\ell}$ has a non-zero VEV, but $\Phi^{(ij)}_q$ and
$\Phi^{(ij)}_{\overline{q}}$ clearly do not, as color is unbroken,
so only the terms in the cyclic permutation that are explicitly
written out in Eq. (6) contribute to quark and lepton masses. The
$\Phi^{(ij)}_{\ell}$ contains, among other components, $H^{(ij)}_u$
and $H^{(ij)}_d$. (See Appendix A.) These have $G_F$-invariant
mass-squared terms of the form

\begin{equation}
M^2_u H^{(ij)*}_u H^{(ij)}_u + M^2_d H^{(ij)*}_d H^{(ij)}_d +
\Delta^2 H_u^{(ij)} H_d^{(ij)} + \Delta^{2*} H_u^{(ij)*}
H_d^{(ij)*}.
\end{equation}

\noindent As already explained in section 2, anthropic fine tuning
will make light a ${\bf 5}$ of light doublets, $H^{(ij)} = \cos
\theta_H H_u^{(ij)} + e^{i \alpha} \sin \theta_H H_d^{(ij)*}$. The
terms in Eq. (7) are $G_F$-invariant, and so leave these 5 light
doublets degenerate. (Their degeneracy will be lifted when $G_F$ is
dynamically broken.) As shown in Appendix A, the terms in Eq. (7)
come from the terms

\begin{equation}
\begin{array}{l}
M_{\Phi}^2 \Phi^{(ij)*}_{\ell}  \Phi^{(ij)}_{\ell}
\\ \\ + [M_{\Delta} \Phi^{(ij)}_{\ell} \Phi^{(ij)}_{\ell}
\Phi_{\ell} + H.c.]
\\ \\ + \sigma Tr ( \Phi^{(ij)*}_{\ell}
\Phi^{(ij)}_{\ell} ) Tr ( \Phi^*_{\ell} \Phi_{\ell}) + \rho Tr
(\Phi^{(ij)*}_{\ell} \Phi^{(ij)}_{\ell} \Phi^*_{\ell} \Phi_{\ell}),
\end{array}
\end{equation}

\noindent where $\Phi = \Phi_q + \Phi_{\overline{q}} + \Phi_{\ell}$
is a singlet under $SO(3)_F$ and gets superlarge VEVs in the
SM-singlet components of $\Phi_{\ell}$. (The traces in Eq. (8) refer
to $G_U$ indices.) Of course, along with the terms in Eq. (8) come
those that result from the $S_3$ permutations.

The sector that dynamically breaks $SO(3)_F$ consists of fermions in
the fundamental and anti-fundamental representations of a confining
gauge group $SU(N)_{DSB}$: $\chi^{\mu i} = (1,3,N)$ and
$\overline{\chi}_{\mu I} = (1,1, \overline{N})$, where $I$ is just a
label runs from 1 to 3. These form condensates, which without loss
of generality can be written $\langle \chi^{\mu i}
\overline{\chi}_{\mu I} \rangle = (f_N)^3 \delta^i_I$ and which
break $SO(3)_F$ completely. There is also a set of several real
$SO(3)_F$ triplet scalars distinguished by the label $J$: $\eta^i_J
= (1,3,1)$. These are ``messenger fields" that communicate the
breaking of $SO(3)_F$ to the Standard Model fields. These have an
explicit positive mass-squared term, with mass that is naturally
superheavy, and also a Yukawa coupling to the $\chi^{\mu i}$,
$\overline{\chi}_{\mu I}$:

\begin{equation}
{\cal L}_{DSB} = \frac{1}{2} (M^2_{\eta})_{JK} \; \eta^i_J \;
\eta^i_K + [y_{IJ}(\chi^{\mu i} \; \overline{\chi}_{\mu I})\;
\eta^i_J + H.c.].
\end{equation}

\noindent Note that we adopt the summation convention that repeated
indices are summed over, for indices of all types, including labels
like $I$, $J$, and $K$ above. When the $\chi^{\mu i}$ and
$\overline{\chi}_{\mu I}$ form a condensate, it gives a linear term
for the $\eta^i_J$ that induces a VEV

\begin{equation}
\langle \eta^i_J \rangle = (f_N)^3 \; (M_{\eta}^2)^{-1}_{\;\;JK} \;
y_{iK}.
\end{equation}

\noindent $M_{\eta}^2$ is naturally superlarge, while the scale
$f_N$ is dynamically generated by the $SU(N)_{DSB}$ interactions and
can naturally be of any magnitude. If $f_N$ is at an intermediate
scale, then the VEV of the $\eta^i_J$ can naturally be near the weak
scale. For example, with $M^2_{\eta} \sim (10^{15} {\rm GeV})^2$ and
$f_N \sim 10^{11}$ GeV, one has $\langle \eta^i_J \rangle \sim 1$
TeV.

Since $f_N$ is the scale of the breaking of the local $SO(3)_F$
family symmetry, the family gauge bosons have mass of an
intermediate scale and produce negligible flavor-changing neutral
current (FCNC) interactions. The gauge symmetries of the model do
not allow any direct renormalizable couplings of the fields
$\chi^{\mu i}$ and $\overline{\chi}_{\mu I}$ to the light fields
(i.e. to the quarks and leptons and light scalar doublets). The
breaking of $SO(3)_F$ is communicated to the light fields by the
$\eta^i_J$. (Note that the $\eta^i_J$ are superheavy, even though
their VEVs are near the weak scale. The group $SO(3)_F$, again, is
broken at high scales by the dynamical condensate.) In particular,
it is easily seen that there is only one renormalizable term that
gives $SO(3)_F$-breaking splittings of the multiplet of light scalar
doublets, namely

\begin{equation}
{\cal L}'_{\Phi} = \lambda_{KI} \left( \Phi^{(ij)*}_{\ell} \;
\Phi^{(jk)}_{\ell} \; \eta^k_K \; \eta^i_I \right).
\end{equation}

\noindent If we define the hermitian matrix $m^2$, which we shall
call the ``master matrix", by  $(m^2)^{ki} \equiv \eta^k_K \;
\lambda_{KI} \; \eta^i_I$ (remembering the summation convention),
then the terms in Eq. (11) give $SO(3)_F$-breaking masses to the 5
light scalar doublets $H^{(ij)}$ of the form

\begin{equation}
H^{(ij)*} H^{(jk)} (m^2)^{ki} = Tr [H^* \; H \; m^2].
\end{equation}

\noindent As explained in section 2, anthropic tuning will set the
mass-squared of the lightest of the five scalar doublets to be
negative and of order -(100 GeV)$^2$. The other four scalar doublets
will then have mass-squared of order $m^2 \sim \langle \eta^i_J
\rangle^2$. Since these other four scalar doublets will mediate FCNC
processes, their masses must be at least several TeV. On the other
hand, they should not be much larger than this, as else they will
not give unification of gauge couplings.

Which linear combination of the five $H^{(ij)}$ is the lightest
(i.e. which linear combination is the SM Higgs dioublet $H_{SM}$)
directly determines the ``textures" of the Yukawa couplings of
$H_{SM}$ to the light quarks and leptons. For example, if $H_{SM}$
were purely $H^{(23)} \subset \Phi^{(23)}_{\ell}$, then by Eqs. (2)
and (6) one sees that there would be terms $y \cos \theta_H \;
(\overline{u}_2 u_3 + \overline{u}_3 u_2) \; H_{SM} + y \; e^{i
\alpha} \; \sin \theta_H (\overline{d}_2 d_3 + \overline{d}_3 d_2)
\; H_{SM} + y' \; e^{i \alpha} \sin \theta_H \; (\ell_2^+ \ell_3^- +
\ell_3^+ \ell_2^-) \; H_{SM}$, so that the textures would all be of
the same form, having only non-vanishing 23 and 32 elements. In
general, however, the master matrix $(m^2)^{ki} = \eta^k_K \;
\lambda_{KI} \; \eta^i_I$ is a non-trivial $3 \times 3$ hermitian
matrix. Therefore, one expects $H_{SM}$ to be a linear combination
of all 5 of the $H^{(ij)}$ and the quark and lepton mass textures to
have all their elements non-zero. In particular, if $H_{SM} =
\sum_{ij} a_{ij} H^{(ij)}$, then the textures will simply be
proportional to $a_{ij}$.

One of the interesting features of this kind of model, therefore, is
that there is a direct connection between the spectrum of the light
scalar doublets and the pattern of Yukawa couplings of the Standard
Model Higgs field. The master matrix determines the pattern of
masses and mixings of the scalar doublets, which in turn determines
which linear combination is the Standard Models Higgs field, which
then in turn determines the quark and lepton textures. A hierarchy
among the splittings of the scalar doublets leads to a hierarchy
among the quark and lepton masses. Suppose, for example, that the
master matrix $m^2$ has the hierarchical form

\begin{equation}
m^2 = m_0^2 \left( \begin{array}{ccc} 1 + \delta_{11} & \delta_{12}
& \delta_{13} \\ \\ \delta_{12} & \delta_{22} & \delta_{23} \\
\\ \delta_{13} & \delta_{23} & \delta_{33} \end{array} \right), \;\;\;
\delta_{ij} \ll 1, \;\; m^2_0 > 0.
\end{equation}

\noindent Then one can see that the term $[H^{(ij)*}  H^{(jk)}
(m^2)^{ki}]$ will give mass-squared contributions to those
$H^{(ij)}$ which have one or more indices equal to 1 (namely
$H^{(12)}$, $H^{(13)}$, and $H'^{(11)} \equiv (2 H^{(11)} - H^{(22)}
- H^{(33)})/\sqrt{6}$) that are of of order $m_0^2$, while it gives
mass-squared contributions to the others ($H^{(23)}$ and $H'^{(22)}
\equiv (H^{(22)} - H^{(33)})/\sqrt{2}$) that are only
$O(\delta_{ij}) \; m_0^2$. The lightest scalar doublet is the
Standard Model Higgs $H_{SM}$, whose mass-squared is pushed slightly
negative by anthropic fine tuning. We will call the next lightest
the ``Lightest Extra Scalar Doublet" $H_{LESD}$. These two lightest
doublets, $H_{SM}$ and $H_{LESD}$, are approximately linear
combinations of $H'^{(22)}$ and $H^{(23)}$ and are therefore only
split from each other by $O(\delta_{ij}) m_0^2$. The other three
scalar doublets are split from these two by $O(m_0^2)$. The pattern
is schematically shown in Fig. 2.

\vspace{0.5cm}

\begin{picture}(360,180)
\put(160,40){\line(1,0){60}} \thicklines
\put(160,38){\line(1,0){40}} \put(160,50){\line(1,0){40}}
\put(160,130){\line(1,0){40}} \put(160,160){\line(1,0){40}}
\put(150,40){\vector(0,1){13}} \put(150,48){\vector(0,-1){13}}
\put(100,40){$O(\delta_{ij}) m^2_0$} \put(120,126){$\frac{1}{2}
m_0^2$} \put(120,156){$\frac{2}{3} m_0^2$}
\put(210,156){$H'^{(11)}$} \put(210,126){$H^{(12)},H^{(13)}$}
\put(210,55){$H_{LESD}$} \put(210,60){\vector(-1,-1){10}}
\put(210,24){$H_{SM}$} \put(210,28){\vector(-1,1){10}}
\put(230,36){$m^2=0$} \put(160,0){{\bf Fig. 2}}
\end{picture}

\vspace{0.2cm}

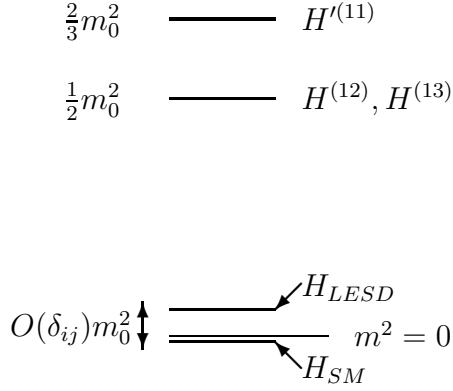
\noindent {\bf Fig. 2.} A schematic plot of the mass-squared
spectrum of the ${\bf 5}$ of scalar doublets in the toy $SO(3)_F$
model.

 \vspace{0.5cm}

\noindent The splittings shown in Fig. 2 come from diagonalizing the
explicit form of the mass matrix of the scalar doublets, which (from
Eq. (13)) is to leading order in $\delta_{ij}$ given by

\begin{equation}
[H'^{(22)} H^{(23)} H'^{(11)} H^{(12)} H^{(13)}] \left[
\begin{array}{ccccc} \frac{\delta_{22} + \delta_{33}}{2} &
\frac{\delta_{23}^*- \delta_{23}}{2} & \frac{\delta_{33} -
\delta_{22}}{2 \sqrt{3}} & \frac{\delta_{12}}{2} & -
\frac{\delta_{13}}{2 \sqrt{2}} \\ &&&& \\
\frac{\delta_{23}- \delta_{23}^*}{2} & \frac{\delta_{22} +
\delta_{33}}{2} & \frac{\delta_{23}^* - \delta_{23}}{2\sqrt{3}} &
\frac{\delta_{13}}{2 \sqrt{2}}
& \frac{\delta_{12}}{2} \\ &&&& \\
\frac{\delta_{33} - \delta_{22}}{2 \sqrt{3}} & \frac{\delta_{23} -
\delta_{23}^*}{2\sqrt{3}} & 2/3 & \frac{2 \delta_{12}^* -
\delta_{12}}{2\sqrt{3}} &
- \frac{\delta_{13}}{2\sqrt{6}} \\
&&&& \\ \frac{\delta_{12}^*}{2} & \frac{\delta_{13}^*}{2\sqrt{2}} &
\frac{2\delta_{12} - \delta_{12}^*}{2\sqrt{3}} & 1/2 &
\frac{\delta_{23}^*}{2} \\ &&&& \\
-\frac{\delta_{13}^*}{2\sqrt{2}} & \frac{\delta_{12}^*}{2} & -
\frac{\delta_{13}^*}{2 \sqrt{6}} & \frac{\delta_{23}}{2} & 1/2
\end{array}
\right] m_0^2 \left[ \begin{array}{c} H'^{(22)} \\ \\ H^{(23)} \\ \\
H'^{(11)} \\ \\ H^{(12)} \\ \\ H^{(13)} \end{array} \right],
\end{equation}

\noindent Therefore, the lightest scalar doublet $H_{SM}$ is
predominantly a linear combination of $H'^{(22)} = (H^{(22)} -
H^{(33)})/\sqrt{2}$ and $H^{(23)}$, with $O(\delta_{ij})$ admixtures
of the others. (Note that this depends on the sign of the largest
elements in $m^2$, which were chosen in Eq. (13) to be positive.)
Specifically, one finds $H_{SM}$ to be of the form

\begin{equation}
\begin{array}{ll}
H_{SM} \cong & \cos \gamma \; ((H^{(22)} - H^{(33)})/\sqrt{2} + \sin
\gamma H^{(23)} \\ \\ & + O(\delta_{12}, \delta_{13}) H^{(12)} +
O(\delta_{12}, \delta_{13}) H^{(13)} \\ \\ & + O((\delta_{22} -
\delta_{33}), Im(\delta_{23})) (2 H^{(11)} - H^{(22)} -
H^{(33)})/\sqrt{6}).
\end{array}
\end{equation}

\noindent That implies that the quark and lepton textures have the
form

\begin{equation}
M_{q,\ell} \sim \left( \begin{array}{ccc} O((\delta_{22} -
\delta_{33}), Im(\delta_{23})) & O(\delta_{12}, \delta_{13}) &
O(\delta_{12}, \delta_{13}) \\ \\ O(\delta_{12}, \delta_{13}) & \cos
\gamma & \sin \gamma \\ \\ O(\delta_{12}, \delta_{13}) & \sin \gamma
& - \cos \gamma \end{array} \right) \langle H_{SM} \rangle.
\end{equation}

\noindent Note that the largest (smallest) elements of these
textures correspond to the lightest (heaviest) scalar doublets.
Moreover, the ratios of splittings within the 5 of scalar doublets
is closely related to the ratios of quark and lepton masses. The
splitting between the two lightest scalar doublets turns out to be
(from Eq. (14)) $O(Im(\delta_{23}), (\delta_{22} - \delta_{33})^2)
m_0^2$, while the splitting that separates these two doublets from
the three heavier doublets is $O(1) m_0^2$. See Fig. 2. Compare this
to the ratio of masses of the first family of fermions to the masses
of the heavier families, which is $O((\delta_{22} - \delta_{33}),
Im(\delta_{23}))$, as can be seen from Eq. (16).

One sees from Eq. (16) that this $SO(3)_F$ model is not realistic,
because the quark and lepton mass matrices have to be traceless.
(The 5 of $SO(3)$ is a traceless tensor.) One cannot therefore have
a threefold fermion mass hierarchy: if one of the families is made
very light, the tracelessness forces the other two families to have
nearly equal and opposite masses to each other.

Another unrealistic feature of this toy model is that the up quark,
down quark, and charged lepton mass matrices (which shall be denoted
$M_U$, $M_D$, $M_L$) are all proportional to the same matrix, namely
$\langle H^{(ij)} \rangle$. Consequently there is no CKM mixing. The
distinction between the different types of fermions (up quarks, down
quarks, and charged leptons) comes from the breaking of the unified
group $SU(3)_c \times SU(3)_L \times SU(3)_R \times S_3$. The
superlarge VEVs that do this breaking must be invariant under
$SO(3)_F$ (otherwise $SO(3)_F$ would be broken at super-large
scales). Therefore, the breaking of $SO(3)_F$, which generates the
non-trivial quark and lepton textures, does not depend on these
$G_U$-breaking VEVs and the textures do not ``know" that the unified
group is broken, and hence the textures have the same form for the
different types of fermions.

A third difficulty of the $SO(3)_F$ model is that there are only
five light scalar doublets, and as Fig. 2 shows, three of them are
orders of magnitude heavier than 1 TeV. (If $O(\delta_{ij}) m_0^2
> 1$ TeV, as must be the case if $H_{LESD}$ does not mediate
excessive FCNC processes, then the other extra light scalar
doublets, whose masses are $O(m_0^2)$, must be several orders of
magnitude heavier than a TeV.) This does not give a good unification
of gauge couplings.

Some of these difficulties can be overcome in the context of
$SO(3)_F$ by making the model more complicated. However, we shall
find in section 4 that they can be overcome in a very simple way by
going to $SO(4)_F$. There will then be four light families; one of
these, however, can be made heavy by ``mating" and getting mass with
an $SO(4)_F$-singlet mirror family. The full mass matrices of the
fermions would then contain a $4\times 4$ block for the four
families in the 4 of $SO(4)_F$. There can be a threefold hierarchy
among the eigenvalues of such a matrix. Tracelessness will then
cause the two largest eigenvalues to be nearly equal and opposite.
That will not matter, however, because one of these large
eigenvalues can be of the family that mates with the mirror family.
Nevertheless, one will find that the near degeneracy of these two
(largest) eigenvalues is related to a near degeneracy of the two
lightest scalar doublets --- just as in the $SO(3)_F$ toy model
(Fig. 2). This is a general feature of this kind of model, and it is
interesting phenomenologically because it means that one of the
``extra" scalar doublets (the LESD) will dominate over all the
others in low-energy phenomenology. That makes these models much
more predictive than they would otherwise be.

Since the LSED is split from the SM Higgs doublet by an amount much
smaller than the other splittings within the $G_F$ multiplet, most
of the ``extra" scalar doublets have to be several orders of
magnitude heavier than a TeV. In an $SO(4)_F$ model this can
compensate for the fact that there are nine scalar doublets, rather
than five or six, and give a good unification of gauge couplings.

\section{A Realistic $SO(4)_F$ Model}

The $SO(4)_F$ model is quite similar to the $SO(3)_F$ toy model
except that in addition to the families in a vector of the family
group, there is a mirror family that is a singlet of the family
group. The quarks and leptons are therefore in two multiplets of
$G_U \times SO(4)_F \times SU(N)_{DSB}$, namely $(F,4,1) = \psi^i =
\psi^i_q + \psi^i_{\overline{q}} + \psi^i_{\ell}$, where $i =
1,...,4$, and $(\overline{F}, 1, 1) = \overline{\psi} =
\overline{\psi_q} + \overline{\psi_{\overline{q}}} +
\overline{\psi_{\ell}}$. The $SO(4)_F$ -singlet mirror family will
mate with one of the families leaving three families light. The
quark and lepton Yukawa terms are (cf. Eq. (6))

\begin{equation}
\begin{array}{ll}
{\cal L}_{Yuk} & = y \; [\psi^i_q \; \psi^j_{\overline{q}}
\Phi^{(ij)}_{\ell} + cyclic] \\ & \\
& + y' \; [\psi^i_{\ell} \; \psi^j_{\ell} \Phi^{(ij)}_{\ell} +
cyclic]
\\ & \\
& + y_J \; [\psi^i_q \overline{\psi_q} \eta^i_J + cyclic],
\end{array}
\end{equation}

\noindent where the $\eta^i_J$ are now in $(1,4,1)$ of the full
gauge group and couple as in Eq. (9) to the $\chi^{\mu i}$,
$\overline{\chi}_{\mu I}$, which are now in $(1,4,N)$ and {\it four}
$(1,1,\overline{N})$.

The quark and lepton textures consequently have the form

\begin{equation}
(f_1 \; f_2 \; f_3 \; f_4 \; \overline{f^c}) \left(
\begin{array}{ccccc} a_f^{11} & a_f^{12} & a_f^{13} & a_f^{14} &
b_f^1 \\ & & & &
\\ a_f^{12} & a_f^{22} & a_f^{23} & a_f^{24} & b_f^2 \\ & & & & \\
a_f^{13} & a_f^{23} & a_f^{33} & a_f^{34} & b_f^3 \\ & & & &
\\ a_f^{14} & a_f^{24} & a_f^{34} & a_f^{44} & b_f^4 \\ & & & & \\
b_{f^c}^1 & b_{f^c}^2 & b_{f^c}^3 & b_{f^c}^4 & 0 \end{array}
\right) \left( \begin{array}{c} f^c_1 \\ \\ f^c_2 \\ \\ f^c_3 \\ \\
f^c_4 \\ \\ \overline{f} \end{array} \right),
\end{equation}

\noindent where $f$ stands for $u$, $d$, or $\ell^-$, and $f^c$
stands for $u^c$, $d^c$, and $\ell^+$. The $4 \times 4$ block is
symmetric and traceless and is given by (see Eqs. (2) and (17))

\begin{equation}
\begin{array}{l}
a^{ij}_u = \cos \theta_H \; y \; \langle H^{(ij)} \rangle, \\ \\
a^{ij}_d = e^{i \alpha} \sin \theta_H \; y \; \langle H^{(ij)} \rangle, \\
\\ a^{ij}_{\ell} = e^{i \alpha} \sin \theta_H \; y' \; \langle
H^{(ij)} \rangle \\ \\
\Longrightarrow a^{ij}_u \propto a^{ij}_d \propto a^{ij}_{\ell}.
\end{array}
\end{equation}

From the third term in Eq. (17) it appears that the entries $b^i_f =
b^i_{f^c} = \Sigma_J \; y_J \; \langle \eta^i_J \rangle$ and that
these are the same for $f= u, d, \ell$. However, as discussed in
Appendix C, there can be $d>4$ effective operators (involving the
superlarge VEVs that break the unification group $G_U$) which have
the effect at low energy of making the Yukawa couplings $y_J$ in Eq.
(17) different for different fermion types. Thus the third term in
Eq. (17) should really be written $y_J^Q \; Q^i \; \overline{Q} \;
\eta^i_J + y^{u^c}_J \;(u^c)^i \; \overline{u^c} \; \eta^i_J +
y^{d^c}_J \;(d^c)^i \; \overline{d^c} \; \eta^i_J + y^L_J \; L^i \;
\overline{L} \; \eta^i_J + y^{\ell^c}_J \; \overline{\ell^c} \;
\eta^i_J$. So that one has $b^i_f = \Sigma_J \; y^f_J \langle
\eta^i_J \rangle$ and $b^i_{f^c} = \Sigma_J \; y^{f^c}_J \langle
\eta^i_J \rangle$, where $f= u,d,\ell^-$, $f^c = u^c, d^c, \ell^+$.
It turns out, as explained in Appendix C, that in simple situations
$y^Q_J = y^{u^c}_J = y^{\ell^c}_J$, but $y^{d^c}_J$ and $y^L_J$ can
be different. (This ultimately stems from the fact that in the
representation $F$ of $G_U$ there are ``extra" superheavy fermions
in each family with the same SM charges as $d^c$ and $L$ and their
conjugates. These extra fermions are called $D^c$ and $L'$ in the
Appendices.) So

\begin{equation}
\begin{array}{ll}
b_u^i & = b_d^i = b_{u^c}^i = b_{\ell^c}^i  \\ \\
& \neq b_{\ell}^i \\ \\
& \neq b_{d^c}^i.
\end{array}
\end{equation}

It follows that the full $5 \times 5$ matrices of the different
fermion types ($u$, $d$, $\ell$) are no longer simply proportional
to each other, and so CKM mixing can occur. Moreover, the
tracelessness of the $a^{(ij)}$ is no longer a problem. For example,
suppose that there is a hierarchy in $a^{(ij)}$ such that its first
and second rows and columns are very small (with the first much
smaller than the second), and suppose that $b^i_u = b^i_{u^c}$
points in the $i = 4$ direction, then the mass matrix of the up
quarks has the form

\begin{equation}
M_U = \left( \begin{array}{ccccc}
a^{11} & a^{12} & a^{13} & a^{14} & 0 \\ &&&& \\
a^{12} & a^{22} & a^{23} & a^{24} & 0 \\ &&&& \\
a^{13} & a^{23} & a^{33} & a^{34} & 0 \\ &&&& \\
a^{14} & a^{24} & a^{34} & -a'^{33} & B \\ & & & & \\
0 & 0 & 0 & B & 0 \end{array} \right),
\end{equation}

\noindent where $a'^{33} = a^{33} + a^{11} + a^{22} \cong a^{33}$.
Since $B \sim \langle \eta \rangle \gg$ TeV, while the $a^{ij}$ are
proportional to the electroweak breaking VEV $\sim 100$ GeV, what
happens with the form in Eq. (21) is that the three observed light
families of up quarks, $u$, $c$, and $t$, are approximately those
with $i = 1,2,3$, while the $i = 4$ up quark gets a mass much
greater than a TeV with the mirror family up quark. The $3 \times 3$
mass matrix of the observed up quarks is then approximately

\begin{equation}
\tilde{M}_U \cong \left( \begin{array}{ccc}
a^{11} & a^{12} & a^{13} \\ & & \\
a^{12} & a^{22} & a^{23} \\ & & \\
a^{13} & a^{23} & a^{33} \end{array} \right),
\end{equation}

\noindent
which is unconstrained by the tracelessness condition
of the $4 \times 4$ matrix $a^{ij}$ and can have a
realistic hierarchy.

As in the $SO(3)_F$ toy model, the hierarchy in the quark and lepton
textures is closely connected to a hierarchy in the spectrum of the
scalar doublets, of which there are 9 in the $SO(4)_F$ model.
Suppose, for example, the master matrix $(m^2)^{ki} \equiv \eta^k_K
\; \lambda_{KI} \; \eta^i_I$ (which is now $4 \times 4$, of course)
has $(m^2)^{11} \gg (m^2)^{22} \gg $ the other elements. Then the
four scalar doublets with an index 1 (namely, $H^{(12)}$,
$H^{(13)}$, $H^{(14)}$, and $H'^{(11)} \equiv (3 H^{(11)} - H^{(22)}
- H^{(33)} - H^{(44)})/\sqrt{12}$) will be much heavier than the
three scalar doublets without an index 1 but with an index 2
(namely, $H^{(23)}$, $H^{(24)}$, and $H'^{(22)} \equiv (2H^{(22)} -
H^{(33)} - H^{(44)})/\sqrt{6}$), which in turn will be much heavier
than the two scalar doublets that have neither a 1 nor a 2 index
(namely $H'^{(33)} \equiv (H^{(33)} - H^{(44)})/\sqrt{2}$ and
$H^{(34)}$). Moreover, these two lightest doublets will have a
relatively small splitting, as shown schematically in in Fig. 3.
(Fig. 3 is not drawn to scale. $M_{unif}$ is supposed to be many of
orders of magnitude larger than $m_0^2$. $H'^{(11)}$ is supposed to
be one or two orders of magnitude heavier than $H'^{(22)}$, and so
forth.)

\vspace{0.5cm}

\begin{picture}(360,180)
\put(160,30){\line(1,0){60}} \thicklines
\put(160,28){\line(1,0){40}} \put(160,40){\line(1,0){40}}
\put(160,60){\line(1,0){40}} \put(160,80){\line(1,0){40}}
\put(160,170){\line(1,0){40}} \put(120,166){$M_{unif}^2$}
\put(120,76){$O(m_0^2)$} \put(225,26){$m^2=0$}
\put(210,76){$H'^{(11)}, H^{(12)}, H^{(13)}, H^{(14)}$}
\put(210,56){$H'^{(22)}, H^{(23)}, H^{(24)}$}
\put(210,45){$H_{LESD}$} \put(210,14){$H_{SM}$}
\put(210,19){\vector(-1,1){10}} \put(210,50){\vector(-1,-1){10}}
\put(180,0){{\bf Fig. 3}}
\end{picture}

\vspace{0.2cm}

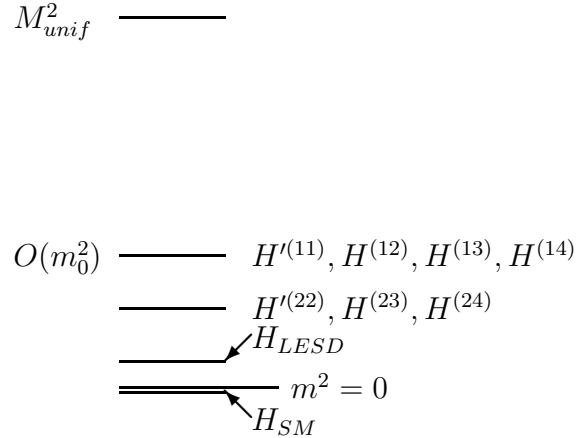
\noindent {\bf Fig. 3.} A schematic plot of the mass-squared
spectrum of the ${\bf 9}$ of scalar doublets in the $SO(4)_F$ model.

\vspace{0.5cm}

\noindent
There is mixing among these scalars, of course, so that the
lightest scalar doublet (the SM Higgs doublet) is predominantly a
linear combination of $(H^{(33)} - H^{(44)})/\sqrt{2}$ and
$H^{(34)}$, but with small admixtures of the others. Corresponding
to this hierarchy, as seen in the last section, the Standard Model
Higgs doublet will have its largest Yukawa couplings in the 33, 34,
43, 44 elements, the next largest in the 22, 23, 32, 24, 42
elements, and the smallest in the 11, 12, 21, 13, 31, 14, 41
elements. Moreover, the ratios of the splittings in the scalar
multiplet are closely related to the ratios of the elements of the
$4 \times 4$ block of the fermion mass matrices.

As explained below, the mass matrices of up quarks, down quarks, and
charged leptons can, without loss of generality, be brought to the
form

\begin{equation}
M_U = \left( \begin{array}{ccccc}
a^{11} & a^{12} & a^{13} & a^{14} & 0 \\ &&&& \\
a^{12} & a^{22} & a^{23} & a^{24} & 0 \\ &&&& \\
a^{13} & a^{23} & a^{33} & a^{34} & 0 \\ &&&& \\
a^{14} & a^{24} & a^{34} & -a'^{33} & B \\ & & & & \\
0 & 0 & 0 & B & 0 \end{array} \right), \end{equation}

\begin{equation}
M_D = t \; \left(
\begin{array}{ccccc}
a^{11} & a^{12} & a^{13} & a^{14} & 0 \\ &&&& \\
a^{12} & a^{22} & a^{23} & a^{24} & 0 \\ &&&& \\
a^{13} & a^{23} & a^{33} & a^{34} & 0 \\ &&&& \\
a^{14} & a^{24} & a^{34} & -a'^{33} & B/t \\ & & & & \\
C_1 & 0 & 0 & C_4 & 0 \end{array} \right), \end{equation}

\begin{equation}
M_L = r \; \left(
\begin{array}{ccccc}
a^{11} & a^{12} & a^{13} & a^{14} & D_1 \\ &&&& \\
a^{12} & a^{22} & a^{23} & a^{24} & D_2 \\ &&&& \\
a^{13} & a^{23} & a^{33} & a^{34} & 0 \\ &&&& \\
a^{14} & a^{24} & a^{34} & -a'^{33} & D_4 \\ & & & & \\
0 & 0 & 0 & B/r & 0 \end{array} \right).
\end{equation}

\noindent These forms are achieved as follows: By the freedom to
choose an $SO(4)_F$ basis, one can simultaneously do the same
$SO(4)_F$ rotation to all the $f^i$ and $(f^c)^i$. Under this, of
course,  the matrices $a^{ij}$ retain their symmetric form and the
vectors $b_u^i = b_d^i = b_{u^c}^i = b_{\ell^c}^i$ can be brought to
the form $(0,0,0,B)$. The vectors $b_{d^c}^i$ and $b_{\ell}^i$,
being different in general, will not be brought to a special form by
this rotation. However, by another change of $SO(4)_F$ basis that
involves rotating only in the 1,2,3 directions, one leaves the form
of $(0,0,0,B)$ unchanged and the vector $b_{d^c}^i$ can be brought
to the form $(C_1,0,0,C_4)$. Finally, by a third cahnge of basis
that involves only the 2,3 directions, one can bring $b^i_{d^c}$ to
the form $(D_1, D_2, 0, D_4)$. The parameters $t$ and $r$ are
defined by $t \equiv \tan \theta_H$, and $r \equiv y'/y$, where $y$
and $y'$ are Yukawa couplings appearing in Eq. (17).

From Eqs. (23)-(25) it can be seen that the effective mass matrices
of the three light families of up quarks, down quarks, and charged
leptons, $\tilde{M}_U$, $\tilde{M}_D$, and $\tilde{M}_L$, depend on
14 parameters: $a^{ij}$, $t$, $r$, $C_1/C_4$, $D_1/D_4$, and
$D_2/D_4$. These must fit 12 observables: six quark masses, three
charged lepton masses, and the three CKM parameters $V_{us}$,
$V_{cb}$, and $V_{ub}$. (The neutrino masses can arise in several
ways, as discussed in Appendix C, and depend on several other
parameters.) If one considers just the quarks, there are 11
parameters to fit 9 quantities. Realistic fits can be obtained, and
will be presented in another place.

The model can fit, but does not predict, the quark and lepton masses
and mixing angles, but by fitting those quantities, one determines
enough parameters of the model to allow one to calculate in terms of
only a few unknown parameters the flavor violation mediated by the
extra scalar doublets --- which in fact is dominated by the exchange
of the lightest extra scalar doublet (LESD), as well as all the
proton-decay branching ratios.

To illustrate how predictive the $SO(4)_F$ model is, consider for
simplicity the case where CP is conserved and all parameters are
real. The traceless part of the $4 \times 4$ symmetric matrix
$(m^2)^{ij}$, which has 9 parameters, determines the complete mass
spectrum of the 9 light scalar doublets (except for the overall mass
of the 9-plet, which is determined anthropically, and is known once
the mass-squared of the SM Higgs doublet is measured directly). That
means that $m^2$ determines which linear combinations of $H^{(ij)}$
the SM Higgs doublet is and therefore the entries $a^{ij}$ in the
mass matrices in Eq. (23). Therefore the 9 parameters in the
``master matrix" $m^2$, together with the two parameters $C_1/C_4$
and $t$ in Eq. (24), determine 8 measureable quantities: the six
quark masses and two CKM mixings. (In this CP conserving case, the
mixing $V_{ub}$ is a real number, and therefore not realistic.)
However, far more than that is also determined. The master matrix
$m^2$ determines the masses of {\it all nine} of the light scalar
doublets and which linear combinations of $H^{(ij)}$ they are.
Consequently, it determines also their Yukawa coupling matrices to
the quarks and therefore all the flavor-changing amplitudes at low
energy, which involves the coefficients of many four-fermion
operators. To put it another way, just fitting the quark masses and
the CKM angles leaves {\it three} undetermined parameters in terms
of which all the FCNC amplitudes in the quark sector can be
calculated. (Actually, there are only two undetermined parameters,
since the unification of gauge couplings gives one constraint on the
mass spectrum of the scalar doublets.)

If one considers also the charged leptons, there is even more
predictivity. Three additional model parameters, namely $r$,
$D_1/D_4$, and $D_2/D_4$, allow one to determine the textures of the
charged leptons, and thus the masses $m_e$, $m_{\mu}$, and
$m_{\tau}$). The net effect, therefore, is that without any more
undetermined parameters being brought in the couplings of the lepton
sector and many more observable quantities can be computed. Among
these are the coefficients of all the flavor-violating four-fermion
operators that involve charged leptons, of which there are many
($\mu_R^+ e_L^- \overline{s}_R d_L$, $\mu_R^- e_L^+ e_R^- e_L^+$,
etc.).

Finally, in terms of just a few more parameters, one can also
predict all the proton decay branching ratios. Proton decay is
mediated by the exchange of $\tilde{D}^{(ij)}$ and
$\tilde{D^c}^{(ij)}$. (Actually, these mix with $\tilde{d}^{(ij)}$
and $\tilde{d^c}^{(ij)}$. See Appendix A.) These are superheavy, and
so the $G_F$-breaking splittings among their masses can be neglected
in computing proton decay. The masses of these colored scalars are
dominated by two terms, which in the notation of Appendix B are
$M'_{\Delta} \; (\langle \tilde{S} \rangle \; \tilde{D^c}^{(ij)} +
\langle \tilde{N^c} \rangle \tilde{d^c}^{(ij)}) \;
\tilde{D}^{(ij)}$. The couplings of these colored scalars are
completely known, since they are part of the same $SO(4)_F$
multiplet with the SM Higgs doublet, and are controlled by the same
Yukawa terms (the first two in Eq. (17)). For example, in the basis
of Eqs. (23)-(25), $\tilde{D}^{(ij)}$ just couples in the $ij(ji)$
direction. There is also proton decay mediated by the $G_F$-singlet
colored scalars $\tilde{D}$, $\tilde{D^c}$. These amplitudes would
depend on an additional parameter. (Again, the pattern of the Yukawa
couplings of these $G_F$-singlet colored scalars is completely
known, since they couple as $\psi^i_q \psi^i_{\overline{q}}
\Phi_{\ell}$ plus similar terms.)

In sum, fitting the quark and charged lepton masses and the CKM
angles, leaves only a small number of undetermined parameters in
terms of which the coefficients of many flavor-violating
four-fermion operators and all the proton branching ratios can be
computed.

The counting is different if complex phases are taken into account.
On the one hand, there are more model parameters (the complex
phases), but most of these can be ``absorbed" by field
redefinitions, and there are also more quantities that in principle
can be measured (the coefficients of CP-violating operators). We
leave the fitting of the quark and lepton masses, and the
predictions of the patterns of FCNC and proton-decay amplitudes to
future work.

\section{Conclusions}

In this paper it is shown how a realistic model can be constructed
in which there is a multiplicity of light scalar doublets (one of
which is the Standard Model Higgs doublet), just as there is a
multiplicity of fermion families. The multiplicity of light scalar
doublets can give satisfactory gauge coupling unification
\cite{sww}. The multiplicity of both light fermions and light
scalars is due to their forming multiplets of a non-abelian family
group. This family group protects the lightness of the ``extra'
scalar doublets by tying their masses to that if the Standard Model
Higgs doublet. The mass of the Standard Model Higgs doublet is
``anthropically tuned" to be small \cite{abds, bk, sweinberg}.

While the anthropic tuning of the scalar masses cannot be tested,
there are many consequences of the non-abelian family symetry that
can be tested. In particular, the couplings of the scalars are all
related to each other by the family symmetry. Knowledge of the quark
and lepton masses and mixings therefore gives much information about
the pattern of couplings of all the scalars. In this way many
predictions of the patterns of the flavor-changing mediated by the
extra light scalar doublets and of proton decay mediated by the
superheavy colored partners of the light scalars can in principle be
extracted. The number of parameters is enormously restricted by the
family symmetry.

Here it has been shown that it is relatively easy to construct
realistic models based on orthogonal family groups, and in
particular one based on $SO(4)_F$ has been described in detail. It
may be possible to use many other kinds of family symmetries, such
as $SU(N)$ or non-abelian discrete symmetries.

The models discussed here are meant to illustrate the utility in
guiding model-building of an extension of the old ``naturalness
principle", which is called here the ``extended naturalness
principle."  This extended principle forbids apparent tunings of
parameters that are not justified {\it either} by symmetry
principles and dynamical mechanisms (as required by the original
naturalness principle), {\it or} by anthropic considerations.
Whereas anthropic tuning of a parameter is not something that can be
directly tested, the requirement that tunings be either
anthropically justified or be ``natural" in the usual sense can
constrain model building and lead to testable scenarios. The kinds
of models presented here, which can be highly predictive, only make
sense (it would seem) in the context of an anthropically tuned
electroweak scale.

\section*{Appendix A}

The group used for unification in this paper is $G_U = SU(3)_c
\times SU(3)_L \times SU(3)_R \times S_3$, where $S_3$ permutes the
three $SU(3)$ factors cyclically. A multiplet that is used both for
a family and for the Higgs field that breaks the electroweak
symmetry is $(3, \overline{3}, 1) + (\overline{3}, 1, 3) + (1, 3,
\overline{3})$, which is denoted $F$ throughout this paper. The
following table shows how Standard Model fields are contained in
this multiplet. Our convention in the Appendices is that unprimed
indices refer to $SU(3)_c$, primed indices refer to $SU(3)_L$,
barred indices refer to $SU(3)_R$, $a=1,2,3$ is a color $SU(3)$
index, $\lambda'$ is a weak $SU(2)$ index, and a tilde over a field
means that it is a boson with the same SM charges as the fermion
field denoted by the same letter. The SM hypercharge is given by
$Y/2 = -\frac{1}{3} \lambda'_8 - \frac{1}{3} \overline{\lambda}_3 +
\overline{\lambda}_8$, where $\lambda_3 = diag (\frac{1}{2}, -
\frac{1}{2}, 0)$ and $\lambda_8 = diag (\frac{1}{2}, \frac{1}{2},
-1)$.

\begin{equation}
\begin{array}{llll}
Rep & Fermions & Bosons & Y/2 \\ & & & \\
(3, \overline{3}, 1) & (\psi_q)^a_{\; \lambda'} = Q = \left(
\begin{array}{c} u \\ d \end{array} \right) &
(\Phi_q)^a_{\; \lambda'} = \tilde{Q} = \left(
\begin{array}{c} \tilde{u} \\ \tilde{d} \end{array}
\right) & 1/6 \\  & (\psi_q)^a_{\; 3'} = D & (\Phi_q)^a_{\; 3'} =
\tilde{D} & -1/3 \\ & & & \\
(\overline{3}, 1, 3) & (\psi_{\overline{q}})^{\overline{1}}_{\; a} =
d^c & (\Phi_{\overline{q}})^{\overline{1}}_{\; a} = \tilde{d^c} & 1/3 \\
& (\psi_{\overline{q}})^{\overline{2}}_{\; a} = u^c &
(\Phi_{\overline{q}})^{\overline{2}}_{\; a}
= \tilde{u^c} & -2/3 \\
& (\psi_{\overline{q}})^{\overline{3}}_{\; a} = D^c & (\Phi_{\;
\overline{q}})^{\overline{3}}_{\; a}
= \tilde{D^c} & 1/3 \\ & & & \\
(1, 3, \overline{3}) & (\psi_{\ell})^{\lambda'}_{\; \overline{1}} =
L' &
(\Phi_{\ell})^{\lambda'}_{\; \overline{1}} = H_d & -1/2 \\
& (\psi_{\ell})^{\lambda'}_{\; \overline{2}} = \overline{L}' &
(\Phi_{\ell})^{\lambda'}_{\; \overline{2}} = H_u & 1/2 \\
& (\psi_{\ell})^{\lambda'}_{\; \overline{3}} = L &
(\Phi_{\ell})^{\lambda'}_{\; \overline{3}} = \tilde{L} & -1/2 \\
& (\psi_{\ell})^{3'}_{\; \overline{1}} = N^c &
(\Phi_{\ell})^{3'}_{\; \overline{1}} = \tilde{N^c} & 0 \\
& (\psi_{\ell})^{3'}_{\; \overline{2}} = e^+ &
(\Phi_{\ell})^{3'}_{\; \overline{2}} = \tilde{e}^+ & 1 \\
& (\psi_{\ell})^{3'}_{\; \overline{3}} = S & (\Phi_{\ell})^{3'}_{\;
\overline{3}} = \tilde{S} & 0 \end{array}
\end{equation}

There are two scalar multiplets that transform as $F$ under $G_U$,
$\Phi^{(ij)}$ and $\Phi$. The former is rank-2 symmetric traceless
tensor under the family group $G_F$, the latter a singlet under
$G_F$. $\Phi^{(ij)} = \Phi^{(ij)}_q + \Phi^{(ij)}_{\overline{q}} +
\Phi^{(ij)}_{\ell}$ and $\Phi = \Phi_q + \Phi_{\overline{q}} +
\Phi_{\ell}$. These couple to the fermion families, which also
transform as $F$ under $G_U$, but as vectors under $G_F$: $\psi^i =
\psi^i_q + \psi^i_{\overline{q}} + \psi^i_{\ell}$.

There are the following types of Yukawa couplings:

\begin{equation}
\begin{array}{l}
\; \; y \; \left[ (\psi^i_q \; \psi^j_{\overline{q}}) \;
\Phi^{(ij)}_{\ell} + (\psi^i_{\overline{q}} \; \psi^j_{\ell}) \;
\Phi^{(ij)}_q + (\psi^i_{\ell} \; \psi^j_q) \;
\Phi^{(ij)}_{\overline{q}} \right] \\ \\
+ y' \; \left[ (\psi^i_q \; \psi^j_q) \; \Phi^{(ij)}_q +
(\psi^i_{\overline{q}} \; \psi^j_{\overline{q}}) \;
\Phi^{(ij)}_{\overline{q}} + (\psi^i_{\ell} \; \psi^j_{\ell}) \;
\Phi^{(ij)}_{\ell} \right] \\ \\
+ Y \; \left[ (\psi^i_q \; \psi^i_{\overline{q}}) \; \Phi_{\ell} +
(\psi^i_{\overline{q}} \; \psi^i_{\ell}) \; \Phi_q + (\psi^i_{\ell}
\; \psi^i_q) \;
\Phi_{\overline{q}} \right] \\ \\
+ Y' \; \left[ (\psi^i_q \; \psi^i_q) \; \Phi_q +
(\psi^i_{\overline{q}} \; \psi^i_{\overline{q}}) \;
\Phi_{\overline{q}} + (\psi^i_{\ell} \; \psi^i_{\ell}) \;
\Phi_{\ell} \right] \end{array}
\end{equation}

\noindent Suppressing the family indices these contain the following
kinds of terms

\begin{equation}
\begin{array}{llll}
& (\psi_q \; \psi_{\overline{q}}) \;
\Phi_{\ell} &
+ (\psi_{\overline{q}} \; \psi_{\ell}) \;
\Phi_q  &
+ (\psi_{\ell} \; \psi_q) \;
\Phi_{\overline{q}} \\ & & & \\
= & (Q \; d^c) \; H_d & + (d^c \; L') \;
\tilde{Q} & + (L' \; Q) \; \tilde{d^c} \\
+ & (Q \; u^c) \; H_u & + (u^c \; \overline{L}')
\; \tilde{Q} & + (\overline{L}' \; Q) \; \tilde{u^c}
\\ + & (Q \; D^c) \; \tilde{L} & + (D^c \; L) \;
\tilde{Q} & + (L \; Q) \; \tilde{D^c} \\
+ & (D \; d^c) \; \tilde{N^c} & + (d^c \; N^c) \;
\tilde{D} & + (N^c \; D) \; \tilde{d^c} \\
+ & (D \; u^c) \; \tilde{e^+} & + (u^c \; e^+) \;
\tilde{D} & + (e^+ \; D) \; \tilde{u^c} \\
+ & (D \; D^c) \; \tilde{S} & + (D^c \; S) \;
\tilde{D} & + (S \; D) \; \tilde{D^c},
\end{array}
\end{equation}

\noindent
and

\begin{equation}
\begin{array}{llll}
& (\psi_{\ell} \; \psi_{\ell}) \;
\Phi_{\ell} &
+ (\psi_q \; \psi_q) \;
\Phi_q  &
+ (\psi_{\overline{q}} \; \psi_{\overline{q}}) \;
\Phi_{\overline{q}} \\ & & & \\
= & (L \; e^+) \; H_d & + (Q \; Q) \; \tilde{D}
& + (d^c \; u^c) \; \tilde{D^c} \\
+ & (L' \; L) \; \tilde{e^+} & + (Q \; D) \;
\tilde{Q} & + (u^c \; D^c) \; \tilde{d^c} \\
+ & (e^+ \; L') \; \tilde{L} & & + (D^c \; d^c)
\; u^c \\
+ & (L \; N^c) \; H_u & & \\
+ & (\overline{L}' \; L) \tilde{N^c} & & \\
+ & (N^c \; \overline{L}) \; \tilde{L} & & \\
+ & (\overline{L}' \; S) \; H_d & & \\
+ & (L' \; \overline{L}') \; \tilde{S} & & \\
+ & (S \; L') \; H_u & & \end{array}
\end{equation}

In the scalar multiplets $\Phi^{(ij)}$ and $\Phi$, only the parts
$\Phi^{(ij)}_{\ell}$ and $\Phi_{\ell}$ have components with non-zero
VEVs, as otherwise color would be broken. $\Phi_{\ell}$ contains
superlarge VEVs in $\tilde{S} \; (= \Phi^{3'}_{\;\; \overline{3}})$
and $\tilde{N^c} \; (= \Phi^{3'}_{\;\; \overline{1}})$, which help
break $G_U$ down to the Standard Model group (which, of course, also
means breaking $S_3$). These large VEVs get rid of the extra
fermions in each family by giving mass to the $D$, $D^c$, $L'$,
$\overline{L}'$, as can be seen from Eqs. (28) and (29). (If
$\langle \tilde{N^c} \rangle = 0$, then $D$ mates purely with $D^c$,
and $\overline{L}'$ mates purely with $L'$. With $\langle
\tilde{N^c} \rangle \neq 0$, however, $D$ mates partly with $d^c$,
so that the light right-handed down quarks are linear combinations
of $D^c$ and $d^c$. Similarly, $\overline{L}'$ mates partly with
$L$, so that the light lepton doublets are linear combinations of
$L$ and $L'$.)

The $\Phi^{(ij)}_{\ell}$ contains the doublets
$H_u^{(ij)}$, $H_d^{(ij)*}$, and $\tilde{L}^{(ij)*}$.
Anthropic fine tuning makes one linear combination
of these light (as discussed below) which we call
$H^{(ij)}$. The lightest of the $H^{(ij)}$ is the
Standard Model Higgs doublet $H_{SM}$.

The $G_F$-invariant masses of the scalar doublets
get contributions from the terms

\begin{equation}
\begin{array}{l}
M_{\Phi}^2 \Phi^{(ij)*}_{\ell}  \Phi^{(ij)}_{\ell}
\\ \\ + [M_{\Delta} \Phi^{(ij)}_{\ell} \Phi^{(ij)}_{\ell}
\Phi_{\ell} + H.c.]
\\ \\ + \sigma Tr ( \Phi^{(ij)*}_{\ell}
\Phi^{(ij)}_{\ell} ) Tr ( \Phi^*_{\ell} \Phi_{\ell}) + \rho Tr
(\Phi^{(ij)*}_{\ell} \Phi^{(ij)}_{\ell} \Phi^*_{\ell} \Phi_{\ell}),
\end{array}
\end{equation}

\noindent where ``{\it Tr}" in the last two terms refers to traces
over the $SU(3)_c \times SU(3)_L \times SU(3)_R$ indices. There are
other terms that are related to those in Eq. (30) by $S_3$
permutations. There are also other quartic terms that do not
contribute to the superlarge masses of the scalar doublets. The
terms in Eq. (28) give a mass-squared matrix for the scalar doublets
of the form

\begin{equation}
(H_u, H_d^*, \tilde{L}^*) \left( \begin{array}{ccc} M^2 & M_{\Delta}
\langle \tilde{S} \rangle &
M_{\Delta} \langle \tilde{N^c} \rangle \\ && \\
M_{\Delta}^* \langle \tilde{S} \rangle^* & M^2 + \rho | \langle
\tilde{N^c} \rangle |^2 & \rho \langle \tilde{S} \rangle^* \langle
\tilde{N^c}
\rangle \\ & & \\
M_{\Delta}^* \langle \tilde{N^c} \rangle^* & \rho \langle
\tilde{N^c} \rangle^* \langle \tilde{S} \rangle  & M^2 + \rho |
\langle \tilde{S} \rangle |^2 \end{array} \right)
\left( \begin{array}{c} H_u^* \\ \\ H_d \\ \\
\tilde{L} \end{array} \right),
\end{equation}

\noindent where $M^2 = M^2_{\Phi} + \sigma (| \langle \tilde{S}
\rangle |^2 + | \langle \tilde{N^c} \rangle |^2 )$.

In the discussion in the main text, only the mixing of $H_u$ and
$H_d$ were considered, not $\tilde{L}$, for ease of discussion.
Including the mixing with $\tilde{L}$ does not qualitatively affect
the conclusions reached in the text. Note that in general there are
three unequal eigenvalues of this matrix. Also, by having two
parameters ``scan", such as $M^2_{\Phi}$ and $M_{\Delta}$, both the
lightest eigenvalue $\mu^2$ and the parameter that was called $\tan
\theta_H$ in the text will scan.

\section*{Appendix B}

From the Yukawa couplings shown in Appendix A one sees that
$\tilde{D}$, $\tilde{D^c}$ can mediate proton decay. The terms $(d^c
\; N^c) \tilde{D}$, $(u^c \; e^+) \tilde{D}$ conserve $B$ and $L$
only if $\tilde{D}$ has $B= \frac{1}{3}$, $L= 1$; whereas the term
$(Q \; Q) \tilde{D}$ conserves $B$ and $L$ only if $\tilde{D}$ has
$B= - \frac{2}{3}$, $L=0$.  Since both kinds of terms are present
$\tilde{D}$ exchange mediates proton decay. Similar arguments apply
to $\tilde{D^c}$. (On the other hand, the exchange of $\tilde{Q}$
does not cause dangerous proton decay. The terms $(d^c \; L')
\tilde{Q}$ and $(D^c \; L) \tilde{Q}$ conserve $B$ and $L$ if
$\tilde{Q}$ has $B = \frac{1}{3}$ and $L= -1$. Then the term $(Q \;
D) \tilde{Q}$ violates $B$ and $L$, but this term contains the
purely superheavy quark $D$, and so does not produce rapid proton
decay.)

The question is whether the $\tilde{D}$ and $\tilde{D^c}$ can be
made heavy enough to avoid rapid proton decay, while leaving $G_U$
unbroken down to the scale $10^{14}$ GeV.  This can be done by the
terms $(M'_{\Delta} \Phi^{(ij)}_q \Phi^{(ij)}_{\overline{q}}
\phi_{\ell} + cyclic)$ and $M^{\prime \prime}_{\Delta} \Phi_q
\Phi_{\overline{q}} \Phi_{\ell} + cyclic)$. These give masses
$(M'_{\Delta} \langle \tilde{S} \rangle) \tilde{D}^{(ij)}
\tilde{D^c}^{(ij)}$ and $(M^{\prime \prime}_{\Delta} \langle
\tilde{S} \rangle) \tilde{D} \tilde{D^c}$. Since $G_U$ invariance
allows $M'_{\Delta}$ and $(M^{\prime \prime}_{\Delta}$ to be
arbitrarily large, the masses of the dangerous colored scalars can
be much larger than $10^{14}$ GeV.  Note that these are different
terms than the $(M_{\Delta} \Phi^{(ij)}_{\ell} \Phi^{(ij)}_{\ell}
\phi_{\ell} + cyclic)$ that are responsible for the $\delta^2$ term
in the scalar doublet mass matrix (cf. Eq. (8) of the text), and
unrelated to it by $S_3$. Making the dangerous scalars heavy to
suppress proton decay does create split multiplets that give
threshold corrections to the running of the gauge couplings above
the scale $10^{14}$ GeV.

\section*{Appendix C}

As discussed after Eq. (19) in the text, the Yukawa couplings $y_J
\; [\psi^i_q \; \overline{\psi_q}^i \; \eta^i_J + cyclic]$ must be
different for the different types of fermions (up quarks, down
quarks, and charged leptons) in order to get realistic mass
matrices. If they are not different, then all the $3 \times 3$ mass
matrices of the observed quarks and leptons, $\tilde{M}_U$,
$\tilde{M}_D$, and $\tilde{M}_L$ will be proportional to each other,
giving no CKM mixing and unrealistic mass relations. To make the
couplings different, the superlarge VEVs that break the unified
group must come into the low energy Yukawa couplings. This can
happen in a simple way if superheavy fields that get mass from these
VEVs are ``integrated out" to give effective $d>4$ Yukawa terms.

Suppose, for example, that in the $SO(4)_F$ model of section 4 there
is an additional superheavy family-mirror family pair, $\psi' +
\overline{\psi}'$ The complete set of quarks and leptons is thus
$(F, 4, 1) = \psi^i$, $(\overline{F}, 1,1) = \overline{\psi}$,
$(F,1,1) = \psi'$, $(\overline{F}, 1,1) = \overline{\psi}'$. Then
the $G_F$-singlet family $\psi'$ will ``mate" with some linear
combination of the two $G_F$-singlet mirror families
$\overline{\psi}$ and $\overline{\psi}'$ to get a superlarge mass,
leaving one mirror family light. Superlarge VEVs that break the
unified group can also contribute to these superlarge masses. Thus,
the mirror family that remains light will consist of linear
combinations of $\overline{\psi}$ and $\overline{\psi}'$ that
``know" about the breaking of the unified group.

Consider the following terms;

\begin{equation}
y_J \; (\psi^i \overline{\psi}) \; \eta^i_J + y'_J \; (\psi^i \;
\overline{\psi}') \; \eta^i_J.
\end{equation}

\noindent Suppose that for a type of fermion $f$, the ``light"
$\overline{f}$ (the one that does not get a superlarge mass) is a
combination $\overline{f} = \alpha_f \; \overline{f}_{(\psi)} +
\beta_f \; \overline{f}_{(\overline{\psi}')}$. Then, the above term
gives

\begin{equation}
[\alpha_f \; (y_J \; \langle \eta^i_J \rangle + \beta_f \;
(y'_J \; \langle \eta^i_J \rangle )] \; f^i \overline{f}.
\end{equation}

\noindent Since, in general, $y_J \; \langle \eta^i_J \rangle$ and
$y'_J \; \langle \eta^i_J \rangle$ point in different directions in
$SO(4)_F$ space, and $\alpha_f$ and $\beta_f$ can be different for
different fermion types $f$, one sees that the desired difference in
the textures can result: the $1 \times 4$ and $4 \times 1$ blocks
can be different in $M_U$, $M_D$, and $M_L$.

The superheavy fermion mass terms that are relevant are

\begin{equation}
\begin{array}{l}
(M_1 \; \psi' \; \overline{\psi} + M_2 \psi' \;
\overline{\psi}')_{[q\; q + cyclic]} \\ \\+ (\lambda_1 \; \psi \;
\psi \Phi + \lambda_2 \psi' \; \psi' \; \Phi + \lambda_3 \psi \;
\psi' \Phi)_{[\ell \; \ell \; \ell + cyclic]} \\ \\+ (\rho_1 \; \psi
\; \psi \Phi + \rho_2 \psi' \; \psi' \; \Phi + \rho_3 \psi \; \psi'
\Phi)_{[q \; \overline{q} \; \ell + cyclic]} \\ \\ + (\lambda_4 \;
\overline{\psi} \; \overline{\psi} \Phi^* + \lambda_5
\overline{\psi}' \; \overline{\psi}' \; \Phi^* + \lambda_6
\overline{\psi} \; \overline{\psi}' \Phi^*)_{[\ell \; \ell \; \ell +
cyclic]} \\ \\ + (\rho_4 \; \overline{\psi} \; \overline{\psi}
\Phi^* + \rho_5 \overline{\psi}' \; \overline{\psi}' \; \Phi^* +
\rho_6 \overline{\psi} \; \overline{\psi}' \Phi^*)_{[q \;
\overline{q} \; \ell + cyclic]}.
\end{array}
\end{equation}

\noindent the complete problem will not be analyzed explicitly here,
but the significant points that emerge from such an analysis will be
indicated. If the superheavy VEVs of $\Phi_{\ell}$ (namely $\langle
\tilde{S} \rangle$ and $\langle \tilde{N^c} \rangle$) are neglected,
one only has the terms $M_1 \; \psi' \; \overline{\psi} + M_2 \psi'
\; \overline{\psi}')_{[q\; q + cyclic]}$, which treat all types of
fermions the same. The distinction between the different fermion
types comes from the superlarge VEVs of $\Phi_{\ell}$, which break
the unified group $G_U$. But these superlarge VEVs give mass only to
the species $D$, $D^c$/$d^c$, $L'$/$L$, and $\overline{L}'$. That is
why it is only the particle types $d^c$ and $L$ that get
distinguished from the others, as stated in Eq. (20).

Neutrino masses can arise in several ways. Perhaps the simplest is
to introduce fermions that are singlets under $G_U$ and vectors
under the family group, to play the role of ``right-handed
neutrinos" in the type-I see-saw mechanism. For example, in the
$SO(4)_F$ model they would transform under $G_U \times SO(4)_F
\times SU(N)_{DYN}$ as $(1,4,1)$. Denote these by $N^i$. Then the
following couplings would be allowed:

\begin{equation}
\begin{array}{l}
{\cal Y}_1 \; [(\psi_{\ell}^i \; N^j) \; \Phi_{\ell}^{*(ij)} + cyclic] \\ \\
+ {\cal Y}_2 \; [(\psi_{\ell}^i \; N^i) \; \Phi_{\ell}^* + cyclic] \\ \\
+ M_N \; (N^i N^i) \\ \\ \gamma \; [(\Phi^*_{\ell}
\Phi^{(ij)}_{\ell})(\Phi^*_{\ell} \Phi^{(ij)}_{\ell}) + cyclic] +
H.c.
\end{array}
\end{equation}

\noindent Integrating out the $N^i$ would give a tree-level
contribution to the light neutrino masses of the type-I form, namely
$M_{\nu} = - M_{Dirac} \; M_R^{-1} \; M_{Dirac}^T$, with
$(M_{Dirac})^{ij} = {\cal Y}_1 \langle H_d^{(ij)} \rangle^*$ and
$(M_R)^{ij} = M_N \; \delta^{ij}$. This by itself would give
unrealistic neutrino masses, as they would have a strong hierarchy
of masses. There would also be loop contributions to the neutrino
masses that went as $(M_{\nu})^{ij} = \delta^{ij} \frac{1}{16 \pi^2}
\; \gamma \; ({\cal Y}_2)^2 \; \langle H_u^{(k \ell)}
\rangle^2/{\cal M}$, where ${\cal M}$ is a combination of superheavy
masses that arises from the momentum integral of the loop. Other
contributions to the light neutrino mass matrix, both tree-level and
from loops are also possible, depending on what fields and couplings
are present at high scales. Thus, such models are not predictive of
neutrino properties.

\end{document}